\documentclass[12pt,A4]{article}
\usepackage{epsfig}
\usepackage{array}
\usepackage{amsmath}
\usepackage{amssymb}
\newcommand{\be}{\begin{equation}}
\newcommand{\ee}{\end{equation}}
\newcommand{\bea}{\begin{eqnarray}}
\newcommand{\eea}{\end{eqnarray}}

\begin{document}

\large
\newpage
\begin{center}
THE NON-RESONANCE SHAKE MECHANISM OF THE NEUTRINOLESS DOUBLE ELECTRONIC CAPTURE

\medskip
  F. F. Karpeshin\footnote{D.I. Mendeleyev Institute for Metrology.},
M. B. Trzhaskovskaya\footnote{National Research Center ``Kurchatov Institute'' --- Petersburg Nuclear Physics Institute.}, L. F. Vitushkin$^{1}$
\bigskip

Abstract
\end{center}

It is generally accepted that double neutrinoless electron capture
is a resonance process. The calculations of the probability of
shaking with the ionization of the electron shell occurring during
the transformation of $^{152}$Gd and $^{164}$Er nuclei are
performed below. These nuclides have the lowest resonance defect
among all known nuclei, being considered as main candidates for
discovering the neutrinoless mode of the transformation.  The
results show predominant contribution of the new mechanism for
most of the candidate nuclei. The value of this amendment rapidly
increases with an increasing  resonance defect. Thus, in
principle, double neutrinoless $e$-capture appears not to be a
resonance process at all.

\newpage

\section{Introduction}
Recently, the Xenon Collaboration reported on the first ever
direct observation of  $2e2\nu$ capture in the $^{124}$Xe nucleus
[1]. This event was a step of paramount importance in searches for
neutrinoless double electron capture. Its investigation is crucial
for testing the Majorana  nature of the neutrino. This process is
traditionally viewed as a resonance one, since no particle is
emitted upon the respective nuclear transformation [2]. Therefore,
it cannot proceed on isolated nuclei even if the energy deposition
is positive, $Q > 0$. The energy-momentum conservation law
requires the transfer of part of energy and momentum to a third
body, and the electron shell of the atom involved plays the role
of this third body. The emerging vacancies are filled via
fluorescence. The energy-momentum conservation law is restored
after the emission of the first photon whose energy includes the
excess quantity $Q$.

The probability for neutrinoless capture is maximal in the
vicinity of the resonance. At the present time, interest is
therefore focused primarily on nuclei characterized by a small
value of  $Q$. In cases where $Q$ is large, a decrease in the
resonance defect is possible upon electron capture from higher
lying shells, such as the $L_1$ and $M_1$ shells, or to excited
states of the nucleus or atom, in which case the process is more
probable than the process of capture to the ground-state level.
For example, the decay process $^{152}$Gd $\to$ $^{152}$Sm
proceeds with a higher probability {\it via} the capture of $KL_1$
electrons to the ground state of the daughter $^{152}$Sm nucleus.
In that case, the resonance defect is $\Delta$ =  0.919 keV. Among
other candidates, the $^{164}$Er $\to$  $^{164}$Dy transformation,
for which $\Delta$ = 6.82 keV, and the $^{180}$W$\to ^{180}$Hf
transformation, for which $\Delta$ = 11.24 keV, are considered as
the most probable ones [3].

In the present study, we propose an alternative, nonresonance,
mechanism of neutrinoless double electron capture. Since it may
come into play irrespective of the value of $\Delta$, the
contribution of this mechanism decreases more slowly with
increasing resonance defect than the traditional
resonance-fluorescent mechanism does. Here, the restoration of the
energy---momentum conservation law occurs owing to electron-shell
ionization caused by the shake effect. Indeed, the $2e$ capture
process is fast in relation to characteristic atomic times.
Therefore, the change in the internal atomic potential because of
the change in the nuclear charge by two units and the
disappearance of two electrons may he viewed as a sudden effect
[2, 4]. Therefore, it may be accompanied by the shake of electron
shells, with the result that one electron is emitted from the
atom. It is this electron that carries away the excess of energy.
Even in the case of the $^{152}$Gd nucleus, which possesses the
minimum resonance defect, the contribution of this new mechanism
within the model considered below increases the decay probability
by about 23\% in relation to the traditional-mechanism
contribution. At the same time, for the case of other candidates,
characterized by higher values of the resonance defect,  its
contribution will prove to be dominant. Physics foundations of the
nonresonance mechanism are outlined in the next section, where the
respective basic relations are derived simultaneously. The results
of the calculations as applied to $^{152}$Gd and $^{164}$Er  are
presented in Section 4. Section 5 is devoted to discussing the
results obtained in the present study.

\section{EXPRESSIONS DESCRIBING
THE PROBABILITY FOR SHAKE IN DOUBLE ELECTRON CAPTURE}

The newly emerging atoms have the atomic number that is smaller by
two units than the atomic number of initial atoms. The former are
produced as neutral atoms in a specific transition state whose
shell is swollen owing to the presence of two holes in the $K$ and
$L_1$ shells [5].

The energy deposition in the process of neutrinoless double
electron capture is determined by the mass difference between the
neutral atoms involved (the initial, $M_1$, and final, $M_2$
ones); that is,  \footnote{Unless otherwise stated, use is made
here of the relativistic system of units, where $\hbar =c=m_e=1$,
$m_e$ being the electron mass.} \be Q=M_1-M_2\,. \ee Even in the
case of the capture of two $K$ electrons, the daughter atom is
neutral but arises in an excited state, featuring two holes in the
place of the captured electrons. This atomic state was called a
swollen state [5]. Upon electron capture from higher lying shells,
the excitation energy of the daughter nucleus becomes higher
appropriately. We denote by $E_A$ the excitation energy of the
daughter nucleus. The resonance defect can then be written as \be
\Delta = Q-E_A \,.  \label{e2} \ee If, in addition, capture occurs
to an excited state of the nucleus at energy $E_N$, this leads,
for quite large values of $Q$, to a further decrease in the
resonance defect; that is, \be
 \Delta = Q-E_A -E_N\,. \label{e3}
\ee In the case being considered, $^{152}$Gd atoms transform into
$^{152}$Sm atoms {\it via} the simultaneous nuclear capture of $K$
and $L_1$ electrons. The single-electron wave functions for the
initial- and final-state atoms are not orthogonal. This gives rise
to various shake processes in the final nucleus via which one or
several electrons go over to an excited state (shake-up) or even
escape to a continuum, the latter leading to the ionization of the
atom (shake-off). Let us examine in more detail the process of the
second type. Suppose that, upon shake-off, the $i$th electron goes
over to a continuum state $f$. We denote by $\psi_i(r)$ the wave
function describing this electron in the gadolinium parent nucleus
and by $\phi_f(r)$ its continuum counterpart in the field of the
singly ionized samarium daughter atom with a hole in the place of
the $i$th electron. The kinetic energy $E$ of the shake-off
electron is determined by setting to zero the resonance defect in
Eq. (2); that is, \be E = \Delta  - I_i \,, \ee where $I_i$ is the
ionization potential for the $i$th electron. In general, states
$i$ and $f$ are not orthogonal since they are eigenfuncfions of
different Hamiltonians. Accordingly, we denote by $V_Z(r)$ and
$V_{Z-2}(r)$  the single-electron potentials in, respectively, the
parent and daughter nuclei and by $\Delta V(r) \equiv  V_Z(r) -
V_{Z-2}(r)$   the sudden change in the potential. The shake
amplitude then assumes the form [6] \be F_{sh} = \langle
\phi_f|\psi_i\rangle \,. \ee The total amplitude can be
represented as the product \be F_{2e}^{sh} = F_{2\beta}F_{sh}, \ee
where, by $F_{2\beta}$,  we have denoted the
double-electron-capture amplitude proper, which leads to the
formation of the transition state of the samarium atom. For the
total width, we accordingly obtain an expression in the form \be
\Gamma_{2\beta}^{sh} = \Gamma_{2\beta} \sum_i N_i  |\langle
\phi_f|\psi_i\rangle|^2 =
  \Gamma_{2\beta} \sum_i N_i |F_{sh}|^2,
\ee where $N_i$ is the occupation number for the $i$th shell.

\section{ FLUORESCENT NEUTRINOLESS
DOUBLE ELECTRON CAPTURE}

Let us now compare expression (7) with its counterpart derived for
the standard resonance mechanism of neutfinoless electron capture,
for example, on the basis of the model considered in [7]. The
latter is obtained by multiplying the width for the formation of
the doorway state, $\Gamma_{2\beta}$, by the Breit---Wigner
resonance factor; that is, \be \Gamma_{2\beta}^{(\gamma)} =
\Gamma_{2\beta} B_W\,, \ee where \be B_W = \frac{\Gamma /
2\pi}{\Delta^2+(\Gamma/2)^2}\,.    \label{BW} \ee In (\ref{BW})
$\Gamma=\Gamma_K+\Gamma_{L_1}$  is the total width of the swollen
state featuring electron holes in the $K$ and $L_1$ shells and
arising upon resonance $2e$ capture. Comparing expressions (7) and
(8), we obtain the relative correction to the decay probability
per unit time in a physically clear form; that is, \be
g\equiv\Gamma_{2\beta}^{\text{(sh)}} / \Gamma_{2\beta}^{(\gamma)}
= \sum_i N_i |\langle \phi_f|\psi_i\rangle  |^2 / B_W \equiv
\sum_i N_i |F_{\text{sh}}|^2 /B_W \,.  \label{rtio} \ee It should
be noted that expression (5) can be recast into an equivalent form
[6]; that is, \be F_{\text{sh}} \approx I_2 = \frac{\langle
\phi_f|\Delta V(r )|\psi_i\rangle}{\Delta}  \,.   \label{sh2} \ee
In principle, shake is possible in any shell whose ionization
energy is less than $Q$, but expression (11) makes it possible to
understand better that it is maximal for $s$-shell electrons.
Indeed, the potential $\Delta V(r)$ is restricted in space to the
region in the vicinity of orbits of hole states formed upon the
capture of respective electrons. Since the electron-capture
probability is the highest for $s$-shell electrons, this is the
region of $K$ and $L$ shells in our case. Accordingly, the
potential $\Delta V(r)$ is maximal in the region of the nucleus.
It decreases uniformly as the distance from the nucleus grows. As
a result, the region where this potential overlaps outer
electrons, especially electrons of high angular momentum, is
substantially smaller than the respective overlap region for inner
electrons, and this leads to a decrease in the shake probability.
It should be noted that, if, in the shake potential $\Delta V(r)$,
the contribution of the field of captured electrons is disregarded
--- that is, if it is assumed that the shake is associated only
with a change of two units in the nuclear charge --- the
contribution of outer electrons would be overestimated.

To conclude this section, we dwell upon the question of the choice
of field between that of the parent nucleus and that of the
daughter nucleus in calculating the resonance energy of the
emitted photon. This question was analyzed in [2, 4]. With
allowance for the classic study of A.B. Migdal on the shake of an
atom in beta decay [6, 8], it was found in [4] that a
mathematically correct method for solving the problem in question
consisted in the expanding the wave functions for the parent and
intermediate atoms in the basis set of eigenfunctions for the
final atom. Therefore, the resonance energy is determined by
levels of the final $^{152}$Sm  atom. A similar conclusion was
drawn in [2]. Within the present approach, the same answer to the
question of energy follows from Eq. (4): it involves the energy of
precisely the final state of the atom. The same applies to the
case where the traditional resonance-fluorescent mechanism is
dominant: the energy-conservation law determines the first-photon
energy on the basis of the balance for the final state of the
atom.

The situation for the wave functions is different: within the
accuracy of the method, one can calculate the matrix elements in
Eq. (11) with the wave functions for either the initial or the
final atom. For the purpose of tests, we performed calculations
according to Eqs. (5) and (11) with invariable wave functions for
the sake of convenience.

\section{ RESULTS OF THE CALCULATIONS}

The calculations on the basis of Eqs. (5) and (10) were performed
in the single-electron approximation by means of the RAINE code
package [9]. The electron wave functions and energies were
calculated by the self-consistent Dirac---Fock method. With the
aim of obtaining deeper insight into underlying physics, we have
calculated the matrix elements in Eqs. (5) and (11) for a number
of hypothetical values of $\Delta$  between 0.05 and 10 keV for
electrons whose ionization potentials are smaller than a preset
value of $\Delta$  and which therefore contribute to the amplitude
for the nonresonance mechanism of the process being considered.

The wave functions used are normalized to unity for discrete
states and in the energy scale for continuum states. Therefore,
the square of the matrix element $F_{\text{sh}}$ has the dimension
of inverse energy. The matrix elements and Breit---Wigner factors
are presented in the relativistic system of units. For the widths
of the hole states, we used the values of $\Gamma_K$ = 20 eV and
$\Gamma_{L_1}$ = 5 eV [10]. By employing the experimental result
of $Q$ = 55.70(18) keV from the most precise measurement reported
in [7], together with the calculated value of $E_A$ = 54.794(9)
keV from the same article, we obtain the resonance defect  of
$\Delta$ =  0.91(19) keV.

The results of the calculations are given in Table I and in Fig.
1. The partial contributions of $s$~electrons from various shells
are quoted in Table 1. The results in question confirm that the
methods of calculations on the basis of Eqs. (5) and (11) are
nearly equivalent. The respective matrix elements are somewhat
different at small values of $Q\approx$  0.5 keV, but, as $\Delta$
grows, this difference decreases to a few percent, starting from
$\Delta$ =  0.7 keV.

The total contribution of the nonresonance mechanism from all
electrons with respect to the resonance mechanism is shown in Fig.
1. The probability for
\begin{figure}[!bt]
%\centerline{\epsfxsize=10cm\epsfbox{RatEn.pdf}}
\includegraphics[width=\textwidth]{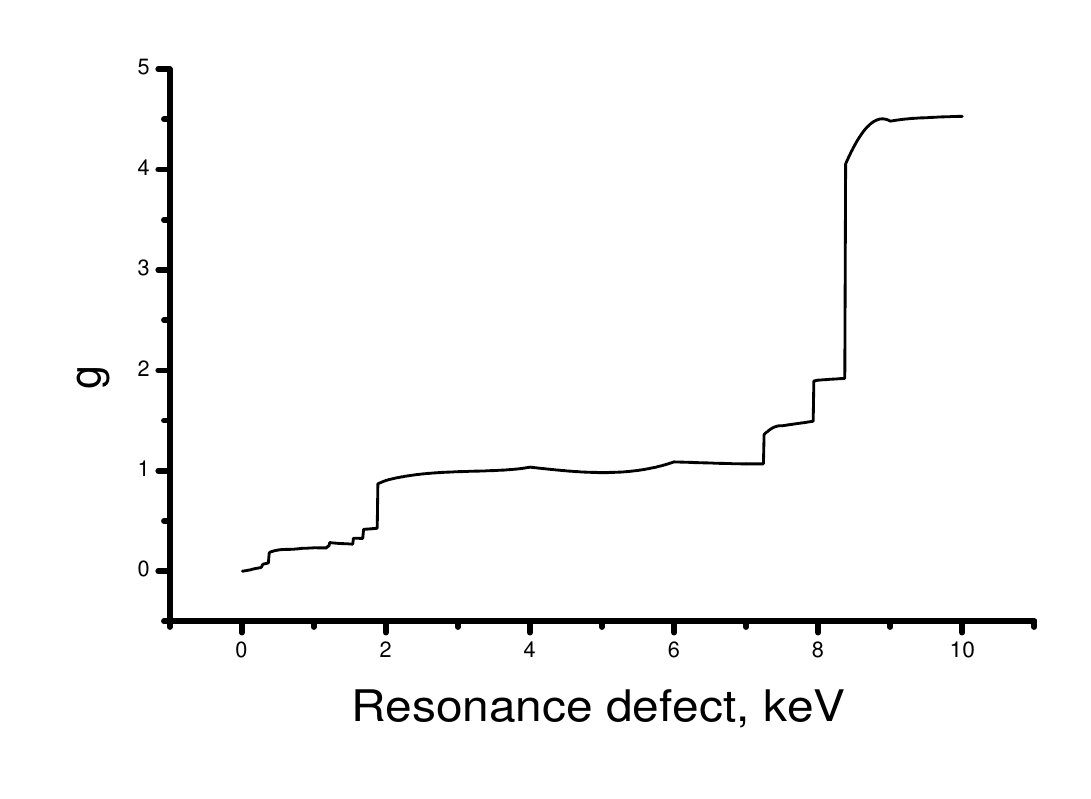}
\caption{\footnotesize { Ratio $g$ (\ref{rtio}) of the
contributions of the nonresonance shake mechanism and the
traditional resonance-fluorescent mechanism of the probability of
neutrinoless double electron capture in $^{152}$Gd versus the
resonance defect~$\Delta$.}} \label{fRatio}
\end{figure}
this process has a manifest stepwise character owing to the fact
that, as $Q$ grows, ever deeper lying shells come into play; it is
noteworthy that the deeper the shell, the greater its contribution
at the threshold. As might have been expected, a dominant
contribution comes from the $4s$-, $5s$- and  $6s$-shell
electrons. The remaining shells --- predominantly the $p_{1/2}$,
$p_{3/2}$ and $d_{3/2}$ ones --- make an additional contribution
of about 30\%. One can see that, at small values of $Q$, the
resonance mechanism is dominant. At the real value of $Q$  =
0.919 keV, the contribution of the nonresonance shake mechanism is
23\%, but, at $Q$ = 3.43 keV, the contributions of the two
mechanisms in question would become identical. If the value of $Q$
were 10 keV, the nonresonance-mechanism contribution would have
been 4.5 times as great as the resonance contribution.
\begin{table}
\caption{Results of calculations for the matrix elements
$F_{\text{sh}}$ in (5) and for the relative shake-effect-induced
correction $g$ in (10) to the probability for neutrinoless
electron capture for $s$-shell electrons (also given here for the
sake of comparison are the values of the sum of the squares of the
matrix elements in Eqs. (5) and (11) --- $\Sigma_{\text{sh}}$ and
$\Sigma_2$, respectively)}
\begin{center}
\begin{tabular}{c||c|c|c|c|c|c|c}
\hline  \hline $\Delta$, keV &
\multicolumn{3}{c|}{$F_{\text{sh}}=\langle \phi_f|\psi_i\rangle $
}&
$\Sigma_ {\text{sh}}$ & $\Sigma_2$ & $B_W$ & $g$ \\
\cline{2-4}
  &$4s$ & $5s$ & $6s$  & $\Sigma_ {\text{sh}}$ & & &  \\
\hline
0.5  & 0.72 & 0.30 & 0.08 & 0.62 & 0.90 & 8.13 & 0.15 \\
0.65 & 0.57 & 0.22 & 0.06 & 0.38 & 0.48 &  4.81 & 0.16 \\
0.83 & 0.45 & 0.17 & 0.05 & 0.23 & 0.26 & 2.95 & 0.16  \\
1.01 & 0.37 & 0.14 & 0.04 & 0.16 & 0.16 & 1.99 & 0.16 \\
1.2 & 0.31 & 0.12 & 0.03 &  0.11 &  0.11 & 1.41 & 0.16 \\
1.5 & 0.25 & 0.09 & 0.02 & 0.071 &   0.064 & 0.90 &  0.16  \\
\hline  \hline
\end{tabular}
\end{center}
\end{table}

    This expectation is fully confirmed in the case of neutrinoless double-electron capture in $^{164}$Er with $Q$ = 6.82(12) keV \cite{eli}. The result of calculation is presented in Fig. 2. The nonresonance mechanism turns out to be by a factor of 3  more effective than the traditional one.
\begin{figure}[!bt]
%\centerline{ \epsfxsize=10cm\epsfbox{ErLL.eps}}
\includegraphics[width=\textwidth]{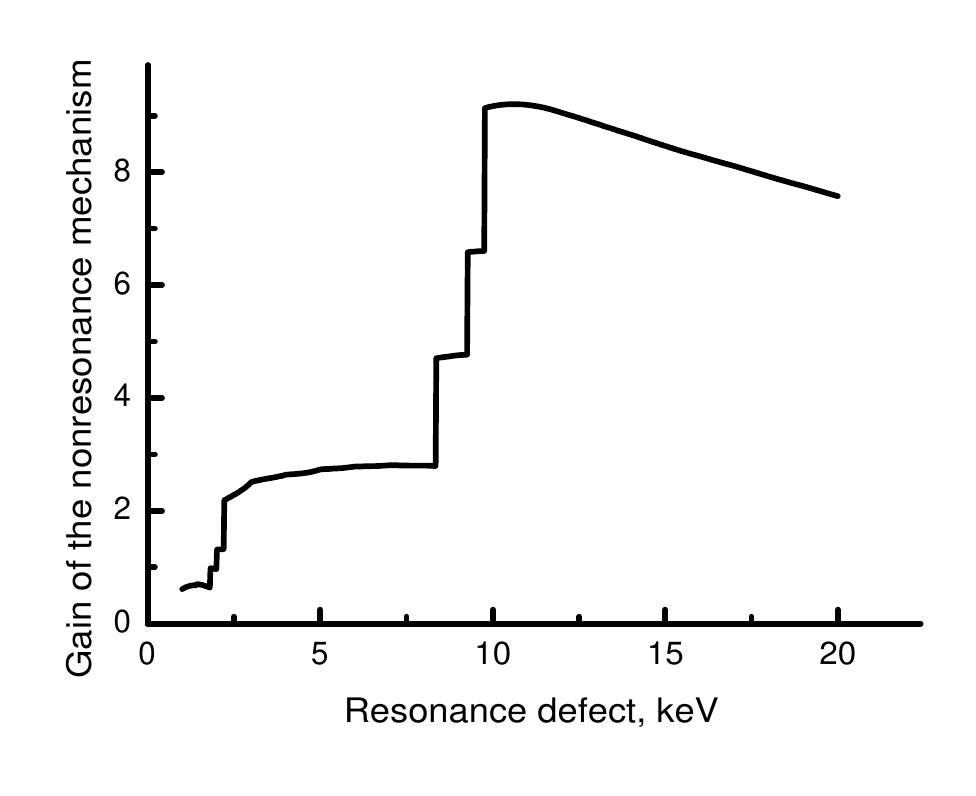}
\caption{\footnotesize {The same ratio as in Fig 1 in the case of
$^{164}$Er $2e0\nu$ capture. At its value of $Q$ = 6.82(12) keV,
the gain attains  3 times. }} \label{Er164LL}
\end{figure}

\section{ DISCUSSION OF THE RESULTS}
We have initiated the study of shake processes in neutrinoless
nuclear double electron capture. As a result, we have proposed a
new mechanism of neutrinoless nuclear double electron capture.
This is a nonresonance mechanism. Therefore, it is natural to
expect that, it will turn out to be more probable for decay in
nuclei characterized by a significant energy deposition. In the
case of the neutrinoless mechanism, such cases may arise in
combination with a high resonance defect entailing a significant
decrease in the resonance-decay probability. An analysis of the
nonresonance mechanism makes it possible to refine substantially
the estimation of the decay half-life. The calculations performed
here for $^{152}$Gd have confirmed this assumption: the inclusion
of the new mechanism leads to an increase of 23\% in the
probability for double capture in relation to the traditional
resonance-fluorescent mechanism. Taking into account the half-life
of 10$^{20}$ yr estimated in [7] for this nucleus with respect to
the $0\nu 2e$ capture mode per effective neutrino mass of
$m_{\beta\beta}$ = 1 eV, we obtain the refined half-life estimate
of $T_{1/2}^{0\nu} \approx 8.1\times 10^{25} \left |\frac{\text{1
eV}}{m_{\beta\beta}}\right |^2$ yr. In
view of the merits of $^{152}$Gd:  as a candidate for measurement of $0\nu 2e$ capture [7], such as a nearly nonexistent background from $2\nu 2e$ capture because of the smallness of the phase space for this process and the largest value of the resonance enhancement factor in (9), this nuclide is one of most probable candidates.  Next candidate is $^{164}$Er. The nonresonance mechanism shortens its lifetime by three times, thus making it also attractive  candidate in searches for neutrinoless double electron capture as an indication of the Majorana nature of the neutrino. \\

\bigskip

\begin{center} ACKNOWLEDGMENTS \end{center}

We are grateful to Yu.N. Novikov for stimulating discussions.
Thanks are also due to I. Alikhanov for enlightening comments.

\end{document}